\documentclass[aps,amsmath,prep,epsfile,nofootinbib,11pt]{revtex4}
\usepackage{graphics}
\usepackage{slashed}
\usepackage{amssymb}
\usepackage{amsmath}
\usepackage{graphicx}

\begin{document}

\title{Radiating Bardeen Black-Hole as a dark matter candidate}
\author{$^{1}$ Alan Sebasti\'an Morales\footnote{alanmorales@mdp.edu.ar}, $^{1,2}$ Juan Ignacio Musmarra\footnote{jmusmarra@mdp.edu.ar}, $^{1,2}$ Mauricio Bellini
\footnote{mbellini@mdp.edu.ar}}
\affiliation{
$^1$ Instituto de Investigaciones F\'{\i}sicas de Mar del Plata (IFIMAR), \\
Consejo Nacional de Investigaciones Cient\'ificas y T\'ecnicas (CONICET), Mar del Plata, Argentina.\\
$^2$ Departamento de F\'{\i}sica, Facultad de Ciencias Exactas y
Naturales, Universidad Nacional de Mar del Plata, Funes 3350, C.P. 7600, Mar del Plata, Argentina.}

\begin{abstract}
We consider a Bardeen black hole (BBH) with variable mass and magnetic charge within the framework of an extended model of gravity and electrodynamics, where the boundary terms in the varied action are treated as geometric quantum fluctuations of spacetime. We propose that the extended Riemann manifold is generated by a new connection, $\hat{\delta\Gamma}^{\mu}_{\alpha\beta}$. These connection components contain geometric information regarding both gravitational and electromagnetic fluctuations on the background curved metric described by a Riemann manifold. Furthermore, we calculate the expectation values for the magnetic field components using a nonlinear and non-commutative quantum algebra for the gravitational and electromagnetic potentials. Our results are consistent with a radiating massive magnetic monopole that produces a radial component of the magnetic field confined to distances $r<\sqrt{3/\lambda_0}$, which are restricted by the vacuum energy density. The field $\hat{A}^{\mu}$ describes massive photons found in the vicinity of a BBH, forming a gravito-electromagnetic structure analogous to that of an atom, which could be a candidate for dark matter in the early universe.
\end{abstract}
\maketitle
\section{Introduction}

While the Schwarzschild and Kerr solutions dominate the study of compact objects, the Bardeen black hole (BBH) offers a unique perspective on the nature of event horizons without the mathematical pathology of infinite density. The BBH \cite{bar} represents a landmark solution in gravitational physics as the first regular black hole candidate. The Bardeen metric is characterized by \cite{rin}
\begin{equation}\label{met}
ds^2=c^2\,f(r) \, dt^2 - \left[\frac{dr^2}{f(r)}  + r^2\,\left( d\vartheta^2+\sin^2{(\vartheta)}\,d\phi^2\right)\right],
\end{equation}
where $c$ is the light velocity in the vacuum and $f(r)$ is the lapse function
\begin{equation}\label{fr}
f(r)= 1- \frac{2\,M\,r^2}{\left(r^2+g_0^2\right)^{3/2}}.
\end{equation}
Here, $M$ is the mass parameter and $g_0$ is a stabilization parameter, which is interpreted in the context of non-linear electrodynamics \cite{gibbons,kali,fati,tosh} as a magnetic monopole charge \cite{beato,gosh}. Unlike the classical Schwarzschild solution, where the Kretschmann scalar $K = R_{\mu\nu\rho\sigma}R^{\mu\nu\rho\sigma}$ diverges as $r \to 0$, the Bardeen geometry ensures that all curvature invariants remain finite throughout the manifold. In this framework, the central singularity is replaced by a de Sitter-like core, where a vacuum-like pressure prevents the final collapse. This model serves as a crucial prototype for studying the phenomenology of quantum-corrected black holes, the shadows of compact objects, and the potential resolution of the information paradox in a semi-classical gravity regime.

The BBH exhibits a de Sitter-like behavior in the limit $r \to 0$, effectively replacing the singular point with a high-density core of finite curvature. This model has sparked extensive research into the thermodynamic properties of regular black holes \cite{Man:2014,boya,buri,bori}, the stability of their Cauchy horizons \cite{Flachi:2013}, and more recently, the observational signatures of their shadows \cite{Abd} in the context of the Event Horizon Telescope (EHT) results. The extension of static regular solutions to include angular momentum is a fundamental step in making these models astrophysically relevant, as rotation is an intrinsic property of most compact objects \cite{Bambi:2013}. The Kerr-Bardeen geometry eliminates the ring-singularity characteristic of the standard Kerr metric, replacing it with a regular core even in the presence of spin. The resulting metric preserves the non-singular nature of the spacetime at $r \to 0$, and introduces significant modifications to the ergosphere. These deviations from classical General Relativity are of particular interest for analyzing the shadows of black holes and the relativistic broadening of $K_{\alpha}$-iron lines in accretion disks.

In this work we will study the partial evaporation of a BBH \cite{hai,berto} as it emits charged massive particle radiation using a recently introduced relativistic formalism, in which such radiation can be geometrically described, such that these particles alter the background geometry during the evaporation of the BBH. To describe the partial evaporation and radiation emission by the BBH, we will consider that the magnetic charge $g_0(\tau)$ and the mass of the BBH, $M(\tau,r)$, evolve over time. The manuscript is organized as follows: in Sect. II we review some introductory mathematical concepts such as variations of connections and the covariant derivation with a self-interaction parameter $\eta$, in order to then be able to calculate the parameter $\lambda(x^{\mu})$, which is related to the vacuum energy from the boundary terms of the varied action. In Sect. III we study
a partially radiating BBH where the mass and the magnetic charge decrease with time, on a background metric that takes into account the fluctuations of the gravitoelectromagnetic field \cite{martin} due to the radiation flux emitted by the BBH, for a static relativistic observer. The remaining background metric is a de Sitter with a $\lambda_0$ that depends on the gauge we choose. We propose a nonlinear quantum algebra that enables us to calculate the dynamic equations for the expectation values of electromagnetic potentials, which are quantized together with gravitational field fluctuations, which are normalized. We obtain a magnetic field solution confined to radii restricted by the vacuum energy, where the $\hat{A}^{\mu}$ field describes massive photons in the vicinity of a BBH, which are also known as dark photons \cite{...,caputo,east}. Consequently, BBHs represent viable dark matter candidates in the early universe. Finally, in Sect. IV we develop some final comments.

\section{Preliminaries}

Standard quantum field theory assumes a fixed background, yet gravity defines the very stage of its own interaction. This inherent self-consistency suggests that the fundamental measure of the universe is not the point-like event, but the global action functional.
The Einstein field equations for the background metric are derived after requiring that the E-H action \footnote{
The notation used along the manuscript is
\begin{itemize}
\item In general, we shall denote with a hat: ${\hat{O}}$, the operatorial character of ${O}$.
\item The partial derivative with respect to some coordinate $x^{\alpha}$ of some quantum scalar field $\hat{\Omega}$ is denoted as $\hat{\partial}_{\alpha}\hat{\Omega}\equiv \hat{\Omega}_{\alpha}$, and therefore $g^{\alpha\beta}\hat{\Omega}_{\alpha}=\hat{\Omega}^{\beta}$.
\item The covariant derivative of some vector field $\hat{\Omega}^{\alpha}$ on a Riemann manifold generated by $g_{\alpha\beta}$, is denoted by $\nabla_{\beta}\hat{\Omega}^{\alpha}$.
    \item The covariant derivative of some vector field $\hat{\Omega}^{\alpha}$ on a Riemann manifold generated by $\bar{g}_{\alpha\beta}$, is denoted by $\bar{\nabla}_{\beta}\hat{\Omega}^{\alpha}$.
\item The Levi-Civita connections related to the components of the metric tensor $\bar{g}_{\alpha\beta}$, are defined as
$\bar{\Gamma}^{\alpha}_{\beta \nu}= \frac{1}{2}\, \bar{g}^{\alpha\mu} \left[\bar{g}_{\mu\beta,\nu} + \bar{g}_{\nu\mu,\beta} - \bar{g}_{\beta\nu,\mu}\right]$.
\item The D'Alambert differential operator on the Riemann manifold generated by ${g}_{\alpha\beta}$, is denoted by ${\Box}\equiv g^{\alpha\beta} \nabla_{\alpha}\nabla_{\beta}$.
\item The D'Alambert differential operator on the Riemann manifold generated by $\bar{g}_{\alpha\beta}$, is denoted by $\bar{\Box}\equiv \bar{g}^{\alpha\beta} \bar{\nabla}_{\alpha} \bar{\nabla}_{\beta}$. In this framework, the Ricci curvature $\bar{R}$, is defined by using the Levi-Civita connections     $\bar{\Gamma}^{\alpha}_{\beta \nu}$.
\item The covariant derivative of $\left( ...\right)$, on the extended manifold, is denoted by $\left( ...\right)_{\| \beta}$ [see Eq. (\ref{uau})].
\item The components of the relativistic velocity when quantum gravitational effects are considered in the action as un external flow with background tensor metric components $g_{\alpha\beta}$, are denoted as $\bar{U}^{\alpha}=\left<B\right|\hat{\Omega}^{\alpha}\left|B\right>$, which are solutions of the geodesic equations: $\frac{d}{dS}\bar{U}^{\alpha} + \Gamma^{\alpha}_{\mu\nu}\,\bar{U}^{\mu}\bar{U}^{\nu} + (z-1)\eta\,\left[\theta(z)-1\right] \,\bar{U}^{\alpha}=0$.
\item The components of the relativistic velocity when quantum gravitational effects are considered in the effective background tensor metric components $\bar{g}_{\alpha\beta}$, are denoted as ${u}^{\alpha}=\left<V\right|\hat{\Omega}^{\alpha}\left|V\right>$, which are solutions of the geodesic equations: $\frac{d}{dS}{u}^{\alpha} + \bar{\Gamma}^{\alpha}_{\mu\nu}\,{u}^{\mu}{u}^{\nu} =0$.
\end{itemize}}
\begin{equation}\label{action}
\mathcal{I} = \int d^4x \sqrt{-g} \left\{ \frac{R}{2\kappa} + \langle B | \hat{\mathcal{L}}_{m} | B \rangle \right\},
\end{equation}
remains constant:
\begin{equation}\label{delta0}
\delta \mathcal{I} = \int d^4 x \sqrt{-g} \left[ \delta g^{\alpha\beta} G_{\alpha\beta} + \kappa \langle B | {\delta g}^{\alpha\beta} \hat{T}_{\alpha\beta} | B \rangle + g^{\alpha\beta} \langle B | \hat{\delta R}_{\alpha\beta} | B \rangle \right] = 0,
\end{equation}
where $\kappa = 8\pi G / c^4$. Here, $\langle B | \hat{\mathcal{L}}_{m} | B \rangle$ represents the background expectation value of the  lagrangian density, $\hat{\mathcal{L}}_{m}$, on the Riemannian manifold. The quantum state $|B\rangle$ is defined in the Heisenberg representation, where quantum operators evolve in spacetime while states are squeezed on the background metric $g_{\alpha\beta}$. Here, $R = g^{\alpha\beta} R_{\alpha\beta}$ is the Ricci scalar, and $g_{\alpha\beta}$ denotes the covariant metric tensor with signature $(+, -, -, -)$ and a vanishing covariant derivative, $\nabla_{\mu}g_{\alpha\beta}=0$. Notice that we have assumed that $\delta g^{\alpha\beta}\,T_{\alpha\beta} ={{\delta g}}^{\alpha\beta}\,\left<B\right|  \hat{T}_{\alpha\beta}\left|B\right>$, where ${{T}}_{\alpha\beta}= 2 \left<B\right|\frac{\delta {\hat{\cal L}}_{m}}{\delta {g}^{\alpha\beta}}\left|B\right>  - g_{\alpha\beta} \left<B\right|{\hat{\cal L}}_{m}\left|B\right>$ is the energy-momentum tensor defined in terms of the Lagrangian density $\hat{\cal L}_{m}$, such that $\left< B |B'\right> =\delta_{BB'} $.

\subsection{Quantum states and Relativistic Ricci flux}

The final terms in \eqref{delta0} represent boundary contributions, which are of quantum gravitational nature:
\begin{equation}\label{RR}
\delta\Theta \equiv \langle B | \hat{\delta\Theta} | B \rangle = \langle B | g^{\alpha\beta} \hat{\delta R}_{\alpha\beta} | B \rangle.
\end{equation}
The quantum background state $\left|B\right>$ must be chosen in a manner that depends on the background metric tensor of the Riemannian manifold
\begin{equation}\label{eq:state}
|B[g_{\mu\nu}; x^{\mu}]\rangle,
\end{equation}
where $g_{\mu\nu}(x^{\mu})$ represents the fundamental classical metric tensor of the background Riemannian manifold, and $x^{\mu}$ denotes the coordinates of spacetime. The dynamics of the quantum state in the Schr\"{o}dinger representation, induced by a local deformation of the spacetime manifold with respect to the line element $S(x^{\mu})=\left[g_{\alpha\beta}\, x^{\alpha}x^{\beta}\right]^{1/2}$ is governed by the equation $i\hbar \,\frac{\hat{\delta}}{{\delta} S} |B[g_{\mu\nu}; x^{\alpha}]\rangle =\bar{U}^{\alpha} \, \hat{P}_{\alpha}\, |B[g_{\mu\nu};x^{\alpha}]\rangle$, that, for $\frac{\hat{\delta}}{{\delta} S}=\bar{U}^{\alpha} \frac{\hat{\delta}}{\delta x^{\alpha}}$, reduces to
\begin{equation}\label{sch1}
i\hbar \frac{\hat{\delta}}{{\delta} x^{\alpha}} |B[g_{\mu\nu}; x^{\alpha}]\rangle = \hat{P}_{\alpha}\, |B[g_{\mu\nu}; x^{\alpha}]\rangle,
\end{equation}
where, $\hat{P}_{\alpha}$ are the quantum momenta related to each coordinate $x^{\alpha}$. In the particular case where the relativistic observer is in a time-like frame and all the spatial components of the relativistic velocities are zero: $\bar{U}^i=0$, the dynamics of the quantum state for $x^a\equiv \tau$ in the Schr\"{o}dinger representation is given by
\begin{equation}\label{sch2}
i\hbar \frac{\hat{\delta}}{{\delta} \tau} |B[g_{\mu\nu}; \tau]\rangle = \hat{H}\, |B[g_{\mu\nu}; \tau]\rangle,
\end{equation}
where $\hat{H}$ is the quantum Hamiltonian.

To integrate the boundary terms into the global dynamics, we posit that the variations in the metric tensor, $\hat{\delta g}_{\alpha\beta}$, serve as the quantum geometric source for $\hat{\delta R}_{\alpha\beta}$. For this assumption to remain consistent within a relativistic framework, we propose a generalized relativistic Ricci flow where $\hat{\delta g}_{\alpha\beta}$ originates from geometric perturbations of the Riemannian manifold induced by a quantum vector field $\hat{\Omega^{\alpha}}$. Following the approach in \cite{ricci1,ricci-flow}, we assume that these quantum variations act as geometrical sources for the varied Ricci tensor operator:
\begin{equation}\label{fff}
\hat{\delta R}_{\alpha\beta} = \lambda(x^{\mu})\, \hat{\delta g}_{\alpha\beta}.
\end{equation}
The components $\delta\Gamma^{\alpha}_{\beta\epsilon} = \langle B | \hat{\delta\Gamma}^{\alpha}_{\beta\epsilon} | B \rangle$ represent the connection's departure from the Levi-Civita connections $\Gamma^{\alpha}_{\beta\epsilon}$. By applying the identity $\delta(g_{\alpha\beta}\,g^{\alpha\beta}) = 0$ and the ansatz \eqref{fff} to the varied action \eqref{delta0}, the boundary terms are effectively assimilated into the modified Einstein field equations:
\begin{equation}
R_{\alpha\beta} - \left[\frac{R}{2}+ \lambda(x^{\mu})\right]\,g_{\alpha\beta} = -\kappa \,T_{\alpha\beta},
\end{equation}
To provide a rigorous description of the extended manifold, we define the quantum connections that quantify the deviation from the classical Riemannian structure. These deviations are treated as geometric manifestations of fundamental quantum gravitational effects \cite{qg1,qg2,qg3,qg4,qg5,qg6}, serving as a framework for the unification of gravity and quantum mechanics. A consistent theory requires the application of covariant derivatives to all quantum tensor fields on this extended geometry, ensuring general covariance. Furthermore, we employ the varied connections to describe the quantum geometric fields $\hat{\Omega}^{\alpha}$ [with expectation value $\left<B\right| \hat{\Omega}^{\alpha}\left| B\right> = \bar{U}^{\alpha}$]:
\begin{equation}\label{uch}
\hat{\delta{\Gamma}}^{\alpha}_{\beta\nu}=b\,\hat{\Omega}^{\alpha}\,g_{\beta\nu},
\end{equation}
which describe the quantum displacement of the extended manifold with respect to the classical Riemannian background defined by the Levi-Civita connections. Here, $\hat{\Omega}^{\alpha}$ is a dimensionless quantum field $4$-vector and $[b]=1/{\rm length}$.

\subsection{Covariant derivatives on the extended manifold}

We shall use the covariant derivative on the extended manifold of a $(n)$-times contravariant and $(m)$-times covariant mixture tensor $\hat{\Upsilon}^{\alpha_1...\alpha_n}_{\beta_1 ... \beta_m}$, which give us the $(n+m+1)$-rank quantum tensor defined in \cite{p2}:
\begin{eqnarray}
&& \hat{\Upsilon}^{\alpha_1...\alpha_n}_{\beta_1 ... \beta_m \|\mu} = \hat{\nabla}_{\mu} \hat{\Upsilon}^{\alpha_1...\alpha_n}_{\beta_1 ... \beta_m} +
\sum_{i=1}^{n} \hat{\delta\Gamma}^{\alpha_i}_{\nu\mu}\,\hat{\Upsilon}^{\alpha_1..\alpha_{i-1}\nu\alpha_{i+1} ..\alpha_n}_{\beta_1 ... \beta_m}
-\sum_{i=1}^{m} \hat{\delta\Gamma}^{\nu}_{\mu \beta_i}\,\hat{\Upsilon}^{\alpha_1...\alpha_n}_{\beta_1 ..\beta_{i-1}\nu\beta_{i+1}. \beta_m}  \nonumber \\
&& -\eta \sum_{i=1}^{n-1} \left(\hat{\Upsilon}^{\alpha_1..\alpha_i\alpha_{i+1}..\alpha_n}_{\beta_1 ... \beta_m}\,\hat{\Omega}_{\mu}
+\hat{\Omega}_{\mu}\,\hat{\Upsilon}^{\alpha_1..\alpha_{i+1}\alpha_i..\alpha_n}_{\beta_1 ... \beta_m}\right) + \eta \sum_{i=1}^{m-1}\left(\hat{\Upsilon}^{\alpha_1...\alpha_n}_{\beta_1 ..\beta_i\beta_{i+1}.. \beta_m}\, \hat{\Omega}_{\mu}+
\hat{\Omega}_{\mu}\,\hat{\Upsilon}^{\alpha_1...\alpha_n}_{\beta_1 ..\beta_{i+1}\beta_i.. \beta_m}\right).\nonumber \\
\label{uau}
\end{eqnarray}
The terms in the last row of (\ref{uau}) describe the interaction of the quantum tensor $\hat{\Upsilon}^{\alpha_1...\alpha_n}_{\beta_1 ... \beta_m}$ with the quantum extended manifold. Furthermore, $\eta$ is a parameter to be determined which is related to the quantum effects and $b=z\eta$ is related to the classical ones. Here, $z$ is a dimensionless parameter which relates $b$ and $\eta$. We define the quantum variation of the quantum tensor field $\hat{\Upsilon}^{\alpha_1...\alpha_n}_{\beta_1 ... \beta_m }$, as
\begin{equation}
\hat{\delta\Upsilon}^{\alpha_1...\alpha_n}_{\beta_1 ... \beta_m}=\hat{\Upsilon}^{\alpha_1...\alpha_n}_{\beta_1 ... \beta_m \|\mu}\,\bar{U}^{\mu},
\end{equation}
where the components $\bar{U}^{\alpha}$ are solutions of the geodesic equations, $\bar{U}^{\epsilon} \left<B\right|\left( \bar{U}^{\alpha}\right)_{\|\epsilon} \left|B\right>=0$:
\begin{equation}\label{geodesic}
\frac{d\bar{U}^{\alpha}}{ds} + \Gamma^{\alpha}_{\mu\nu}\,\bar{U}^{\mu}\bar{U}^{\nu} + (z-1)\eta\,\left[\theta(z)-1\right] \,\bar{U}^{\alpha}=0,
\end{equation}
with $\theta(z)= \frac{\left[3z-4+\sqrt{9\,z^2-8\,z}\right]}{4\,(z-1)}$, $\left<B\right|\hat{\Omega}^{\alpha}\left|B\right>=\bar{U}^{\alpha}$ and $g_{\alpha\beta}\,\bar{U}^{\alpha}\bar{U}^{\beta}=\theta(z)-1$.

We shall define the varied Ricci tensor as an extension of the Palatini prescription \cite{pal}:
\begin{equation}
\hat{\delta R}_{\alpha\beta}= b\,\left[\left(\hat{ \delta\Gamma}^{\mu}_{\alpha\mu} \right)_{\| \beta} - \left(\hat{ \delta\Gamma}^{\mu}_{\alpha\beta}\right)_{\| \mu}\right],
\end{equation}
in the relativistic Ricci flow (\ref{fff}) \cite{ricci1,ricci-flow}.

The covariant derivative of the background metric tensor on the extended manifold, with self-interactions included, is
\begin{equation}\label{gg}
\hat{g}_{\beta\alpha\|\nu} =\nabla_{\nu} g_{\beta\alpha} -b\, \left(g_{\beta\nu} \hat{\Omega}_{\alpha} + \hat{\Omega}_{\beta} g_{\alpha\nu} \right) + 2\eta\, \hat{\Omega}_{\nu} \, g_{\alpha\beta},
\end{equation}
where $\nabla_{\nu} g_{\beta\alpha}=0$ is the covariant derivative of the metric tensor, calculated on the Riemann manifold. Therefore, we obtain the  expressions $\hat{\delta g}_{\alpha\beta}  = b^{-1}\,\left(\hat{g}_{\alpha\beta}\right)_{\|\mu}\,\bar{U}^{\mu}$, and $\hat{\delta g}^{\alpha\beta} = b^{-1}\,\left(\hat{g}^{\alpha\beta}\right)_{\|\mu}\,\bar{U}^{\mu}$, such that we can define
$\bar{g}_{\alpha\beta}  =  {g}_{\alpha\beta}+\left<B\right|\hat{\delta g}_{\alpha\beta}\left|B\right>$.

\subsection{Metric tensor with quantum fluctuations included}

The covariant and contravariant metric tensor components with quantum gravitational corrections are \cite{p2}
\begin{equation}\label{ggg}
\bar{g}_{\alpha\beta}=g_{\alpha\beta}\left[1+\frac{2\left[\theta(z)-1\right]}{z} \right] -2\,\bar{U}_{\alpha}\bar{U}_{\beta}, \qquad
\bar{g}^{\alpha\beta} = \frac{\left[\bar{g}_{\alpha\beta}\right]^{\dagger}}{\left|\bar{g}\right|},
\end{equation}
where $\left[\bar{g}_{\alpha\beta}\right]^{\dagger}$ and $|\bar{g}|$ are, respectively, the adjoint and the determinant of $\bar{g}_{\alpha\beta}$; this construction guarantees the trace identity $\bar{g}_{\alpha\beta}\,\bar{g}^{\alpha\beta}=4$, as required for the inverse of a tensor in a four-dimensional manifold. The components $u^{\alpha}$ denote the new relativistic velocity associated with the background metric $\bar{g}_{\alpha\beta}$; they are solutions of
\begin{equation}\label{geodesica}
\frac{d{u}^{\alpha}}{ds} + \bar{\Gamma}^{\alpha}_{\mu\nu}\,{u}^{\mu}{u}^{\nu} =0,
\end{equation}
such that $\bar{\Gamma}^{\alpha}_{\beta\nu}= \frac{1}{2}\, \bar{g}^{\alpha\mu} \left[\bar{g}_{\mu\beta,\nu} + \bar{g}_{\nu\mu,\beta} - \bar{g}_{\beta\nu,\mu}\right]$.  The new relativistic velocities calculated from the new background metric $\bar{g}_{\alpha\beta}$ are normalized with
$\bar{g}_{\alpha\beta}\,{u}^{\alpha}\,{u}^{\beta}=1$. This metric includes the quantum gravitational fluctuations considered in the boundary conditions of the varied action (\ref{delta0}).

\subsection{Calculation of $\bar{\lambda}(x^{\mu})$ with $\bar{g}_{\alpha\beta}$}

The new fundamental tensor metric with components (\ref{ggg}) incorporates the quantum gravitational effects of the background dynamics. Therefore, we will write a new action with the background metric (\ref{ggg}), on which we define the new background quantum states denoted as $\left|V\right>$:
\begin{equation}\label{I}
\bar{{\cal I}} = \int\,d^4x\,\sqrt{-\bar{g}}\,\left\{\frac{\bar{R}}{2\kappa}+\left<V\right|\hat{\cal L}_m\left|V\right>\right\}.
\end{equation}
Here, $\bar{\lambda}(x^{\nu})$, a parameter calculated on the new Riemann manifold defined by the metric tensor components (\ref{ggg}) [see \cite{p1,pp}], is given by
\begin{equation}\label{la1}
\bar{\lambda}(x^{\nu}) = \frac{3z^2\,\eta}{2} \left[\frac{\eta\,(2z+1)\,\left<V\right|\hat{\Omega}_{\alpha} \hat{\Omega}^{\alpha}\left|V\right> + \left<V\right|\left(\bar{\nabla}_{\beta}\,\hat{\Omega}^{\beta}\right)\left|V\right>}{ (z-4) \left[\theta(z)-1\right]}\right],
\end{equation}
where $\left<V\right|\hat{\Omega}_{\alpha} \hat{\Omega}^{\alpha}\left|V\right>=\left[\bar{\theta}(z)-1\right] $, with
$\bar{\theta}(z)=\frac{4z^3-3z^2-10z+8+(2-z)\,\sqrt{9z^2-8z}}{z(z-1)\left[\sqrt{9z^2-8z}-z\right]}$ \cite{p1}.
The parameter $\bar{\lambda}(x^{\nu})$ has a purely quantum gravitational origin, as a consequence of the flow of $\hat{\Omega}^{\alpha}$ on the Riemann manifold described by $\bar{g}_{\alpha\beta}$. We can choose in (\ref{la1}), the following gauge:
\begin{equation}\label{gauge}
\left<V\right|\bar{\nabla}_{\beta}\,\hat{\Omega}^{\beta}\left|V\right>=C.
\end{equation}
If we define the parameter
\begin{equation}\label{CC}
\lambda_1=\frac{3z^2\,\eta\,C}{2\,\,(z-4)\left[\theta(z)-1\right]},
\end{equation}
we obtain the resulting parameter which takes into account the vacuum energy for the choice of the gauge (\ref{gauge}):
\begin{equation}\label{barlambda}
{\lambda}_0= \frac{3z^2\,\eta}{2} \left[\frac{\eta\,(2z+1)\,\left[\bar{\theta}(z)-1\right]}{ (z-4) \left[\theta(z)-1\right]}\right]+\lambda_1,
\end{equation}
for $z\neq 4$, $z\neq 0$ and $z\neq 1$. Notice that, if we impose the gauge $\left<V\right|\bar{\nabla}_{\beta}\,\hat{\Omega}^{\beta}\left|V\right>=C$, the constant ${\lambda}_0$ can be assimilated in the action (\ref{I}), as well as in the new Einstein equations
\begin{equation}\label{eins}
\bar{R}_{\alpha\beta} - \bar{g}_{\alpha\beta}\left[\frac{1}{2}\, \bar{R}\, + {\lambda}_0\right]= -\kappa\,\left<V\right|\hat{\bar{T}}_{\mu\nu}\left|V\right>,
\end{equation}
such that $\bar{G}_{\mu\nu}=\bar{R}_{\mu\nu}-\left[(1/2)\,\bar{R}+{\lambda}_0\right]\,\bar{g}_{\mu\nu} $ is the Einstein tensor with the cosmological constant ${\lambda}_0$ included, for $\bar{\nabla}_{\alpha}\bar{g}_{\mu\nu}=0$, where now the covariant derivative is defined with respect to the Levi-Civita connections $\overline{\Gamma}^{\alpha}_{\beta \, \nu}= \frac{1}{2}\, \bar{g}^{\alpha\mu} \left[\bar{g}_{\mu\beta,\nu} + \bar{g}_{\nu\mu,\beta} - \bar{g}_{\beta\nu,\mu}\right]$.

The stress tensor $\bar{T}_{\mu\nu}$ is defined as
\begin{equation}
\hat{\bar{T}}_{\mu\nu}= 2 \frac{\delta \hat{\cal L}_m}{\delta \bar{g}^{\mu\nu}} -  \bar{g}_{\mu\nu}\,   \hat{\cal L}_m.
\end{equation}
Furthermore, the components of the relativistic velocity: $u^{\alpha}=\frac{dx^{\alpha}}{\hskip -.15cm ds}$, comply with the normalization $\bar{g}_{\alpha\beta} \,u^{\alpha}u^{\beta}=1$, and are solutions of the geodesic equations
\begin{equation}\label{ge}
\frac{du^{\alpha}}{ds} + \bar{\Gamma}^{\alpha}_{\mu\nu}\, u^{\mu}u^{\nu}=0.
\end{equation}

\section{Partial evaporation of BBH}

To describe the relativistic dynamics of the BBH evaporation, we must solve the geodesic equations for the relativistic velocity components (\ref{geodesic}). To do so, we shall consider a BBH with variable mass $M(\tau,r)$ and magnetic charge $g_0(\tau)$. A particular solution of the static metric (\ref{met}), with (\ref{fr}), is given by
\begin{equation}\label{solu}
g_0(\tau)= \bar{g}_0\,\left[1+e^{-A(z)\,\tau}\right], \qquad M(\tau,r)=\frac{{\lambda}_0}{6}\,\left[r^2+g_0^2(\tau)\right]^{3/2},
\end{equation}
where $\bar{g}_0$ is a constant magnetic charge, $\tau = c\,t$, $A(z)= \eta (z-1)\,\left[\theta(z)-1\right]$ and ${\lambda}_0$ is given by (\ref{barlambda}), for the gauge $\left<V\right|\bar{\nabla}_{\alpha} \Omega^{\alpha}\left|V\right>=C$. Notice that although the mass and charge of the BBH evolve over time, the remnant metric describes a static de Sitter spacetime where the initial magnetic charge of the BBH is given by $\left.{g}_0\right|_{\tau=0}=2\bar{g}_0$. Additionally, we shall require that $\left<V\right|\hat{A}^{\alpha}\left|V\right>=0$. For the particular evolution of $g_0(\tau)$ and $M(\tau,r)$ in (\ref{solu}), we obtain the relativistic solution for (\ref{geodesic}), given by
\begin{equation}
\bar{U}_0 = \frac{{\lambda_0}\, r }{(z-1)\,\eta  \left[3-{\lambda}_0\,r^2\right]},
\end{equation}
with ${\lambda}_0$ given by (\ref{barlambda}). This means that the new Riemann manifold will be generated by the metric tensor components
\begin{eqnarray}
\bar{g}_{00} &=& c^2\,F(r)\,\left[1+\frac{2\left[\theta(z)-1\right]}{z} \right] -2\,\left(\bar{U}_{0}\right)^2, \quad \bar{g}_{11}=
-\frac{1}{F(r)}\,\left[1+\frac{2\left[\theta(z)-1\right]}{z} \right]  , \label{gbar1} \\
\bar{g}_{22} &=& -r^2\,\left[1+\frac{2\left[\theta(z)-1\right]}{z} \right] ,  \qquad \qquad \quad \,\,\bar{g}_{33} = -r^2\,\sin^2{(\vartheta)}\,\left[1+\frac{2\left[\theta(z)-1\right]}{z} \right], \label{gbar2}
\end{eqnarray}
with $F(r)$ given by
\begin{equation}
 F(r)= 1- \frac{{\lambda}_0}{3}\,r^2,
\end{equation}
that corresponds to an effective metric for a de Sitter spacetime. This means that, for a time-like relativistic frame with $u^i=0$, the nonzero covariant and contravariant components for the relativistic velocity will be
\begin{equation}\label{u0}
u_0(r)=\sqrt{\bar{g}_{00}}, \qquad u^0(r)=\frac{1}{\sqrt{\bar{g}_{00}}},
\end{equation}
such that $\bar{g}_{00}\,(u^0)^2=1$.

\subsection{Gravitoelectromagnetic quantum field dynamics}

As in a previous work \cite{apj24}, we shall consider the field $\hat{\Omega}^{\alpha}$ that incorporates both, the geometric gravitational and electromagnetic fluctuations in a unified description
\begin{equation}\label{omega}
\hat{\Omega}^{\alpha} \equiv \hat{\sigma}^{\alpha} + a\,\hat{ A}^{\alpha},
\end{equation}
where $a$ is a coupling constant with units $\left[a\right]={\rm \frac{s \,Coul}{kg\,m}}$, and $\hat{\sigma}$ is the quantum gravitational scalar field such that $\hat{\sigma}_{\alpha} \equiv \hat{\partial}_{\alpha} \hat{\sigma}$. Furthermore, the components $\hat{{A}}^{\alpha}$ are the fluctuations of the electromagnetic $4$-potential. These quantum fields satisfy the dynamics \cite{apj24}
\begin{subequations}
    \begin{equation}
    \bar{ \nabla}_{\mu} \hat{\sigma}^{\mu}+ (2b+\eta) \hat{\sigma}_{\mu}\,\hat{\sigma}^{\mu} +2(2b+\eta)\,a\,\hat{A}_{\mu}
\,\hat{\sigma}^{\mu} -\frac{2}{3}(b-4\eta) \,{\lambda}_0\,{u}_{\mu}\, \hat{\sigma}^{\mu}=0, \label{18}
    \end{equation}
    \begin{equation}
    \bar{ \nabla}_{\mu} \hat{A}^{\mu} + (2b+\eta) \,a\, \hat{A}^{\mu} \hat{A}_{\mu} = \frac{1}{3}(2b-8\eta) \,{\lambda}_0\,{u}_{\mu}\hat{ A}^{\mu}, \label{19}
    \end{equation}
\end{subequations}
where ${\lambda}_0$ is given by the expression (\ref{barlambda}). Notice that the third term in (\ref{18}) describes the coupling between
the gravitational and electromagnetic components of $\hat{\Omega}^{\alpha}$ in Eq. (\ref{omega}).

To describe the background electromagnetic (EM) fields in the absence of external electric sources, we shall consider in the action (\ref{I}), with the Lagrangian density $\hat{\cal L}_{m}$
\begin{equation}\label{lag}
\hat{\cal L}_{m} = -\frac{1}{4} \,\hat{F}^{\mu\nu}\,\hat{F}_{\mu\nu}.
\end{equation}
Here, the electromagnetic tensor $\hat{F}_{\mu\nu}$ is defined by $\hat{F}_{\mu\nu}= \bar{\nabla}_{\mu} \hat{A}_{\nu} - \bar{\nabla}_{\nu} \hat{A}_{\mu}$,
where $\hat{A}^{\mu} \equiv \left( \hat\Phi(\vec{x},\tau)/c, \,\hat{A}^i(\vec{x},\tau)\right)$ is 4-electromagnetic potential such that $\hat{\Phi}(\vec{x},\tau)$ is the electric potential and $\hat{A}^i(\vec{x},\tau)$ is the vector potential. The expectation value for the electromagnetic equations of motion comes from the field equation $\left<V\right|\bar{\nabla}_{\mu} \hat{F}^{\mu\nu}\left|V\right>=0$, that after taking into account the flow field equation (\ref{19}), takes the form
\begin{eqnarray}
\left<V\right|\bar{\Box} \hat{ A}^{\alpha} + \bar{g}^{\alpha\nu}\,\bar{R}_{\nu\mu}\,\hat{ A}^{\mu}\left|V\right> = \left<V\right|\frac{\eta}{3}[2(z-4)]\,\bar{\nabla}^{\alpha}\left({\lambda}_0\,{u}_{\mu}\,\hat{ A}^{\mu}\right)
- \eta (2z+1)\,a\,\bar{\nabla}^{\alpha} \left( \hat{ A}^{\mu} \hat{ A}_{\mu}\right)\left|V\right>,\label{wea}
\end{eqnarray}
where we have considered the fact that for a Riemann manifold without torsion we must require that: $\left<V\right|\bar{\nabla}_{\mu} \bar{\nabla}_{\nu}  \hat{A}^{\alpha} - \bar{\nabla}_{\nu}\bar{\nabla}_{\mu}  \hat{A}^{\alpha}\left|V\right>=
\left<V\right|\bar{R}^{\alpha} _{\hskip.1cm\epsilon\mu\nu} \hat{A}^{\epsilon}\left|V\right>$, with a background Ricci tensor given by  $\bar{R}_{\epsilon\mu}=\bar{R}^{\nu}_{\hskip.1cm\epsilon\mu\nu}=\left.\bar{\Gamma}^{\nu}_{\epsilon\nu}\right._{,\mu}-
\left.\bar{\Gamma}^{\nu}_{\epsilon\mu}\right._{,\nu} + \bar{\Gamma}^{\nu}_{\mu\theta} \bar{\Gamma}^{\theta}_{\nu\epsilon}-
 \bar{\Gamma}^{\theta}_{\theta\nu}\,\bar{\Gamma}^{\nu}_{\epsilon\mu}$ and Levi-Civita connections $\bar{\Gamma}^{\alpha}_{\beta\nu}=\frac{1}{2} \bar{g}^{\alpha\mu} \left[\bar{g}_{\mu\beta,\nu} + \bar{g}_{\nu\mu,\beta} - \bar{g}_{\beta\nu,\mu}\right]$.
Notice that $u^{\nu}$ are solutions of (\ref{geodesica}) and the last term on the right-hand side of (\ref{wea}) is an external current that cannot be attributed to an external magnetic monopole source. Therefore, it can be interpreted as arising from the radiation of a magnetic monopole charge of the BBH. Notice that the quantum state $\left|V\right>$ is in the Heisenberg representation in which quantum operators evolve in spacetime, while the quantum states are squeezed in space and time. Notice that $\left|B\right>$ acts on the background described by $g_{\alpha\beta}$, but the state $\left|V\right>$ acts on the new background with quantum fluctuation included, which is described by $\bar{g}_{\alpha\beta}$. Therefore, the quantum background state $\left|V\right>$ must be chosen in a manner that depends on the new background metric tensor of the Riemannian manifold: \begin{equation}\label{state1}
|V[\bar{g}_{\mu\nu}; \tau]\rangle,
\end{equation}
where $\bar{g}_{\mu\nu}$, given by the components (\ref{gbar1}) and (\ref{gbar2}), represents the fundamental classical metric tensor of the new background Riemannian manifold with quantum gravitational corrections included. The dynamics of the quantum state for a relativistic observer that moves in a time-like frame in the Schr\"{o}dinger representation, is governed by
\begin{equation}\label{eq:schrodinger_corrected}
i\hbar \frac{\hat{\delta}}{{\delta} \tau} |V[\bar{g}_{\mu\nu}; \tau]\rangle = \hat{H}\, |V[\bar{g}_{\mu\nu}; \tau]\rangle,
\end{equation}
where $\hat{H}$ is the quantum Hamiltonian operator. In the case of the BBH these components are given by (\ref{gbar1}) and (\ref{gbar2}).

Let us consider the electromagnetic fields orthogonal to the components of the relativistic velocity ${u}^{\mu}=\frac{dx^{\mu}}{ds}$, so that ${u}^{\mu} \,\left<V\right|\hat{E}_{\mu}\left|V\right>=0$ and ${u}^{\mu}\,\left<V\right|\hat{B}_{\mu}\left|V\right>=0$. The electric and magnetic fields can be written as \cite{review}:
\begin{eqnarray}
\left<V\right|\hat{ E}^{\mu}\left|V\right> &=&\left<V\right| \hat{F}^{\mu\nu}\left|V\right>\,{u}_{\nu},  \label{E} \\
\left<V\right|\hat{ B}^{\mu}\left|V\right> &=&- \frac{1}{2\,\sqrt{-\bar{g}}} \,\epsilon^{\mu\nu\rho\lambda}\, {u}_{\nu}\, \left<V\right|\hat{F}_{\rho\lambda}\left|V\right>. \label{B}
\end{eqnarray}
Here, $\epsilon^{\mu\nu\rho\lambda}$ are the Levi-Civita symbols [the $\sqrt{-\bar{g}}$ appears because we are working in a curved background generated by the metric $(\bar{g}_{\alpha\beta})$ defined in (\ref{gbar1}) and (\ref{gbar2})], such that $\epsilon^{0123}=1,\; (-1)$ for any even (odd) permutation of (0,1,2,3) and 0 if any index is repeated.

\subsection{Quantum fields quantization and quantum algebra}

The equal-time canonical quantization requires that
\begin{equation}
\left[\hat{ A}_{\nu}(\tau,\vec{r}),\,\frac{\delta {\cal \hat{L}}_{m}}{\delta [\bar{\nabla}_{\beta}\hat{ A}_{\alpha}]}(\tau,\vec{r'})\right]= i \,\hbar\,\delta^{\alpha}_{\,\,\,\nu} \,\hat{\sigma}^{\beta}(\tau,\vec{r}) \,\delta^{(3)}\left(\vec{r}-\vec{r}\,'\right), \label{quanti}
\end{equation}
where $\hat{\sigma}^{\beta}\equiv \bar{g}^{\alpha\beta} \,\partial_{\alpha} \hat{\sigma}$ and $\frac{\delta {\cal \hat{L}}_{m}}{\delta [\bar{\nabla}_{\beta}\hat{ A}_{\alpha}]}= \bar{g}^{\mu\beta} \bar{\nabla}_{\mu} \hat{ A}^{\alpha} - \bar{g}^{\alpha\mu} \bar{\nabla}_{\mu} \hat{ A}^{\beta}$. The expression (\ref{quanti}) takes into account the fact that the BBH is not an isolated magnetic charge; it also has mass, and therefore the gravitational field must also be considered in the quantization of the electromagnetic field. In other words, the electromagnetic and gravitational fields form a single entity, and therefore these fields cannot be separately quantized in the BBH. We can define the invariant scalar
\begin{equation}
    \hat{\phi}(x^{\mu},x'^{\mu}) \equiv \hat{A}_{\nu}(x^{\mu})\hat{A}^{\nu}(x'^{\mu}).
\end{equation}
In order to solve the expectation value for the equations (\ref{wea}), we must propose that the following nonlinear quantum algebra for $\hat{A}^{\alpha}$:
\begin{eqnarray}
&& \left< V\right|\hat\partial_{\alpha} \hat{A}^{\beta}({x'}^{\mu})\left|V\right> =  b\,a\,\left<V\right| \hat{A}_{\alpha}({x}^{\mu})\hat{A}^{\beta}({x'}^{\mu})\left|V\right>, \label{qa1} \\
&& \left< V \right|\bar\nabla^{\alpha}\left( \hat{A}^{\mu}({x}'^{\mu})\hat{A}_{\mu}({x}''^{\mu})\right)\left|V\right>
= b\,a^2\,\bar{g}^{\alpha\beta} u_{\beta} \left<V\right| \hat{\phi}(x^{\mu},{x}''^{\mu})\,\hat{\phi}({x}^{\mu},{x'}^{\mu})\left|V\right>, \label{qa2} \\
&& \left<V\right|\hat{\partial}_{\beta}\hat{\sigma}({x'}^{\mu})\left|V\right> = b^2\,u_{\beta} \,\left<V\right|\hat{\sigma}(x^{\mu})\,\hat{\sigma}({x'}^{\mu})\left|V\right>, \label{qa3}
\end{eqnarray}
where, for instance, in (\ref{qa1}), the partial derivative $\hat\partial_{\alpha}$ must be understood as an operator applied to $\hat{A}^{\beta}$. In this framework, the expectation value for the covariant derivative of the quantum electromagnetic potential $\hat{A}^{\beta}$ in Eq. (\ref{qa1}) must be understood as $\left<V\right|\bar{\nabla}_{\alpha} \hat{A}^{\beta}\left|V\right>=
 \left<V\right|\hat\partial_{\alpha} \hat{A}^{\beta}\left|V\right> + \bar{\Gamma}^{\beta}_{\nu\alpha} \left<V\right|\hat{A}^{\nu}\left|V\right>$, for $\left<V\right|\hat{A}^{\nu}\left|V\right>=0$.

 \subsection{Fourier expansion of $\langle\hat{\partial}_{\beta}\hat{A}_{\alpha}\rangle$}

The gauge field $\hat{A}^{\alpha}(x^\mu)$ is quantized covariantly by expanding it in a basis of mode functions $\Phi_{\vec{k}}(x^\mu)$:
\begin{equation}
    \hat{A}^{\alpha}(x^\mu)=\frac{1}{(2\pi)^{3/2}}\sum_{\lambda=0}^{3}\int d^{3}{\vec{k}}
    \left[
        \hat{A}^{\alpha}_{{\vec{k}},\lambda}(x^\mu)
        +
        \left(\hat{A}^{\alpha}_{{\vec{k}},\lambda}(x^\mu)\right)^{\!\dagger}
    \right],
    \label{eq:modeexpansion}
\end{equation}
with $\lambda = 0,1,2,3$ labeling the four polarization states. Each mode can be factorized into a polarization vector, a ladder operator, and a spacetime-dependent mode function, according to
\begin{equation}
    \hat{A}^{\alpha}_{{\vec{k}},\lambda}(x^\mu)
    =
    e^{\alpha\,\lambda}\,\hat{C}_{{\vec{k}},\lambda}\,\Phi_{\vec{k}}(x^\mu),
    \qquad
    \left(\hat{A}^{\alpha}_{{\vec{k}},\lambda}\right)^{\!\dagger}(x^\mu)
    =
    \left(e^{\alpha\,\lambda}\right)^{*}
    \hat{C}^{\dagger}_{{\vec{k}},\lambda}\,\Phi^{*}_{\vec{k}}(x^\mu),
    \label{eq:modefactorization}
\end{equation}
where $e^\alpha_\lambda$ is the polarization four-vector for mode $\lambda$, $\hat{C}_{{\vec{k}},\lambda}$ and $\hat{C}^{\dagger}_{{\vec{k}},\lambda}$ are the annihilation and creation operators for quanta of momentum $\vec{k}$, and $\Phi_{\vec{k}}(x^\mu)$ are scalar functions introduced through the factorization~\eqref{eq:modefactorization} to separate the operator content from the spacetime dependence. In spherical coordinates the operators $\hat{C}_{\vec{k}}=\hat C_{klm}$ and $\hat{C}^{\dagger}_{\vec{k}}=\hat C_{klm}^{\dagger}$, comply with the algebra
\begin{eqnarray}\label{m55}
&& \left[\hat C_{klm},\;\hat C_{k'l'm'}^{\dagger}\right]= \frac{(2l+1)}{4\pi} \frac{(l-m)!}{(l+m)!} \;\delta ({k}-{k'})\, \delta_{ll'}\,\delta_{mm'},\\
&& \left[\hat C_{klm},\;\hat C_{k'l'm'}\right]=
\left[\hat C_{klm}^{\dagger},\;\hat C_{k'l'm'}^{\dagger}\right]=0,
\end{eqnarray}
for $\hat C_{klm}= \hat C_k\, \sqrt{\frac{(2l+1)}{4\pi} \frac{(l-m)!}{(l+m)!}}$. Furthermore, the polarization vectors satisfy
\begin{equation}
    e^{\alpha\,\lambda}\,e_{\alpha}^{\,\,\,\lambda'} = \,\bar{g}^{\lambda\lambda'}, \qquad  e^{0\,\lambda}\,e_{i}^{\,\,\,\lambda'} = \delta^0_{\,\,i}
    \,\bar{g}^{\lambda\lambda'} =0,
    \label{eq:polcomplete}
\end{equation}
so that the four states form an orthonormal basis in the metric signature $(+,-,-,-)$ used in this work. The physical degrees of freedom are selected by the constraint equation~(\ref{19}).

The ladder operators obey
\begin{equation}
    \left[\hat{C}_{{\vec{k}},\lambda},\,\hat{C}_{{\vec{k}'},\lambda'}^{\dagger}\right]
    =
    2\omega_{\vec{k}}\,(2\pi)^{3}\,\bar{g}_{\lambda\lambda'}\,
    \delta^{(3)}(\vec{k}-\vec{k}').
    \label{eq:CCRladder}
\end{equation}
The factor $2\omega_{\vec{k}}$ arises from the requirement of a Lorentz-invariant measure $d^3\vec{k}/(2\omega_{\vec{k}})$ on the mass shell, for the adopted signature: $(+,-,-,-)$. The canonical momentum $\hat{\Pi}_\nu$ follows from the Lagrangian~\eqref{lag}, and satisfies
\begin{eqnarray}
    \left<V\right|\hat{\Pi}_{\nu}(x^{\mu},x'^{\mu})\left|V\right>
    &=& \left<V\right|\hat{\partial}_{0}\hat{A}_{\nu}(x'^{\mu})\left|V\right>  -
    \left<V\right|\hat{\partial}_{\nu}\hat{A}_{0}(x'^{\mu})\left|V\right> \nonumber \\
    & = &
    b\,a\,\left[\left<V\right|\hat{A}_{0}(x^{\mu})\hat{A}_{\nu}(x'^{\mu})\left|V\right>  -
    \left<V\right|\hat{A}_{\nu}(x^{\mu})\hat{A}_{0}(x'^{\mu})\left|V\right>\right],      \label{eq:canonmom}
\end{eqnarray}
where we have made use of the quantum algebra (\ref{qa1}). The vacuum expectation values of the field bilinears are obtained by inserting~\eqref{eq:modeexpansion} and applying $\hat{C}_{\vec{k},\lambda}|V\rangle = 0$:
\begin{align}
    \langle\hat{\partial}_{\beta}\hat{A}_{\alpha}\rangle
    &\sim
    \sum_{\lambda,\lambda'}
    \frac{1}{(2\pi)^{3}}
    \int d^{3}\vec{k} \int d^{3}\vec{k}'\,
    \langle
        (\hat{A}_{\beta})_{\vec{k},\lambda}(x^{\mu})
        (\hat{A}_{\alpha})^{\dagger}_{\vec{k}',\lambda'}(x'^{\mu})
    \rangle
    \notag \\
    &=
    \sum_{\lambda,\lambda'}
    \frac{1}{(2\pi)^{3}}
    \int d^{3}\vec{k} \int d^{3}\vec{k}'\,
    e_{\beta}^{\lambda}\,e_{\alpha}^{\lambda'}\,
    \langle\hat{C}_{\vec{k},\lambda}\hat{C}^{\dagger}_{\vec{k}',\lambda'}\rangle\,
    \Phi_{\vec{k}}(x^{\mu})\,\Phi_{\vec{k}}^{*}(x'^{\mu})
    \notag \\
    &=
    \sum_{\lambda,\lambda'}
    \int d^{3}\vec{k}\;
    2\omega_{\vec{k}}\,
    e_{\beta}^{\lambda}\,e_{\alpha}^{\lambda'}\,
    \bar{g}_{\lambda\lambda'}\,
    \Phi_{\vec{k}}(x^{\mu})\,\Phi_{\vec{k}}^{*}(x'^{\mu}).
    \label{eq:vev_bilinear}
\end{align}
In particular, the expectation value for the equal-time two-point correlation of $\hat{A}_{\mu}(x^{\beta})
\hat{A}_{\nu}(x^{\beta})$ takes the form
\begin{equation}
\langle V|
\hat{A}_{\mu}(\tau,\vec{r})
\hat{A}_{\nu}(\tau,\vec{r}\,')
|V\rangle
=
2\,\sum_{\lambda,\lambda'}
\int d^{3}{\vec{k}}\,
\omega_{\vec{k}}\,
e_{\mu}^{\lambda}\,
e_{\nu}^{\lambda'}\,
\bar{g}_{\lambda\lambda'}\,
\Phi_{\vec{k}}(\tau,\vec{r})\,\Phi^*_{\vec{k}}(\tau,\vec{r}\,'),
\label{eq:correlator_modes}
\end{equation}
so that $\langle V|
\hat{A}_{2}(\tau,\vec{r})
\hat{A}_3(\tau,\vec{r}\,')
|V\rangle$ and $\langle V|\hat{A}_{3}(\tau,\vec{r})\hat{A}_{3}(\tau,\vec{r}\,')|V\rangle$, which are the correlators that enter $B^1$ through Eq.~\eqref{br}, are determined by the spectral distribution of
$\Phi_{k}(\tau,\vec{r})\,\Phi^*_{k}(\tau,\vec{r}\,')$ in the electromagnetic modes on the new background metric $\bar{g}_{\alpha\beta}$.

\subsection{Calculation of  $\left<V\right|\hat{E}^i\left|V\right>$ and $ \left<V\right|\hat{B}^i\left|V\right>$}

Therefore, the dynamics for the expectation value of the fields will be given by Eq. (\ref{wea}) with the algebraic conditions (\ref{qa1}), (\ref{qa2}) and (\ref{qa3}):
\begin{eqnarray}
 & \bar{g}^{\mu\nu}& \left[\bar{\Gamma}^{\alpha}_{\theta\mu} \left<V\right|\hat{A}_{\nu}(x^{\mu})\,\hat{A}^{\theta}(x'^{\mu})\left|V\right> + \bar{\Gamma}^{\alpha}_{\theta\nu} \left<V\right|\hat{A}_{\mu}(x^{\mu})\,\hat{A}^{\theta}(x'^{\mu})\left|V\right> - \bar{\Gamma}^{\epsilon}_{\mu\nu}\left<V\right|\hat{A}_{\epsilon}(x^{\mu})\,
\hat{A}^{\alpha}(x'^{\mu})\left|V\right> \right]\nonumber \\
&=& \frac{2\eta}{3} \,(z-4) \,{\lambda}_0\,u_{\mu} \,\bar{g}^{\alpha\epsilon} \left<V\right| \hat{A}_{\epsilon}(x^{\mu})
\hat{A}^{\mu}(x'^{\mu})\left|V\right>
- \eta\,(2z+1)\,a\,u^{\alpha}\,\left<V\right|\hat{\phi}(x^{\mu})\,\hat{\phi}(x'^{\mu})\left|V\right>.\label{dyn}
\end{eqnarray}
For the case of the BBH, we shall require that
\begin{eqnarray}
    \left<V\right| \hat{A}_{1}(x^{\mu})\hat{A}^{1}(x'^{\mu})\left|V\right>&=&0,\label{pep} \\
    \left<V\right| \hat{A}_{0}(x^{\mu})\hat{A}^{i}(x'^{\mu})\left|V\right>&=&0.\label{pep1}
\end{eqnarray}
The equation in (\ref{pep}) implies that the field $ \hat{A}^{\mu}(x^{\mu})$ propagates in the $1$-direction, and therefore the polarizations of $e^{\lambda}_1=0$. Because of this we must require the additional conditions:
\begin{equation}
\left<V\right| \hat{A}_{1}(x^{\mu})\hat{A}^{i}(x'^{\mu})\left|V\right>=0.
\end{equation}
The expression (\ref{pep1}) can be obtained from (\ref{eq:vev_bilinear}) with the second condition of (\ref{eq:polcomplete}). Here, the first condition reflects the absence of radial electromagnetic radiation, consistent with the spherical symmetry of the BBH background, while the second decouples the temporal component of the gauge potential from its spatial counterparts. Evaluating Eq.~(\ref{dyn}) for each value of the free index $\alpha$ and using the diagonal structure of $\bar{g}_{\mu\nu}$ together with the conditions~(\ref{pep}), one obtains a closed algebraic system for $\langle V|\hat{A}_{\mu}(x^{\mu})\hat{A}^{\nu}(x'^{\mu})|V\rangle$. The Eq.~(\ref{dyn}) for each $\alpha$-component gives us the following set of equations:
\begin{subequations}
\label{system}
\begin{align}
  &(2\bar{g}^{11}\bar{\Gamma}^{0}_{01}-K(r)) \left\langle V \right|\hat{A}_{1}(x^{\mu})\hat{A}^{0}(x'^{\mu})\left| V \right\rangle
   + 2\bar{g}^{00}\bar{\Gamma}^{0}_{10} \left\langle V \right|\hat{A}_{0}(x^{\mu})\hat{A}^{1}(x'^{\mu})\left| V \right\rangle \nonumber \\
  &\quad - \bar{g}^{33}\bar{\Gamma}^{2}_{33} \left\langle V \right|\hat{A}_{2}(x^{\mu})\hat{A}^{0}(x'^{\mu})\left| V \right\rangle \nonumber \\
  &\qquad = \frac{\eta}{3}(2z-8)\lambda_{0}u_{0}\bar{g}^{00} \left\langle V \right|\hat{A}_{0}(x^{\mu})\hat{A}^{0}(x'^{\mu})\left| V \right\rangle
   - \eta(2z+1)a u^{0} \left\langle V \right|\hat{\phi}(x^{\mu})\hat{\phi}(x'^{\mu})\left| V \right\rangle, \label{eq:alpha0} \\[12pt]
  &(2\bar{g}^{11}\bar{\Gamma}^{1}_{11}-K(r)) \left\langle V \right|\hat{A}_{1}(x^{\mu})\hat{A}^{1}(x'^{\mu})\left| V \right\rangle
   + 2\bar{g}^{00}\bar{\Gamma}^{1}_{00} \left\langle V \right|\hat{A}_{0}(x^{\mu})\hat{A}^{0}(x'^{\mu})\left| V \right\rangle \nonumber \\
  &\quad + 2\bar{g}^{22}\bar{\Gamma}^{1}_{22} \left\langle V \right|\hat{A}_{2}(x^{\mu})\hat{A}^{2}(x'^{\mu})\left| V \right\rangle
   + 2\bar{g}^{33}\bar{\Gamma}^{1}_{33} \left\langle V \right|\hat{A}_{3}(x^{\mu})\hat{A}^{3}(x'^{\mu})\left| V \right\rangle \nonumber \\
  &\quad - \bar{g}^{33}\bar{\Gamma}^{2}_{33} \left\langle V \right|\hat{A}_{2}(x^{\mu})\hat{A}^{1}(x'^{\mu})\left| V \right\rangle
   = \frac{\eta}{3}(2z-8)\lambda_{0}u_{0}\bar{g}^{11} \left\langle V \right|\hat{A}_{1}(x^{\mu})\hat{A}^{0}(x'^{\mu})\left| V \right\rangle, \label{eq:alpha1} \\[12pt]
  &(2\bar{g}^{11}\bar{\Gamma}^{2}_{21}-K(r)) \left\langle V \right|\hat{A}_{1}(x^{\mu})\hat{A}^{2}(x'^{\mu})\left| V \right\rangle
   + 2\bar{g}^{22}\bar{\Gamma}^{2}_{12} \left\langle V \right|\hat{A}_{2}(x^{\mu})\hat{A}^{1}(x'^{\mu})\left| V \right\rangle \nonumber \\
  &\quad - \bar{g}^{33}\bar{\Gamma}^{2}_{33} \left\langle V \right|\hat{A}_{2}(x^{\mu})\hat{A}^{2}(x'^{\mu})\left| V \right\rangle
   + 2\bar{g}^{33}\bar{\Gamma}^{2}_{33} \left\langle V \right|\hat{A}_{3}(x^{\mu})\hat{A}^{3}(x'^{\mu})\left| V \right\rangle \nonumber \\
  &\qquad = \frac{\eta}{3}(2z-8)\lambda_{0}u_{0}\bar{g}^{22} \left\langle V \right|\hat{A}_{2}(x^{\mu})\hat{A}^{0}(x'^{\mu})\left| V \right\rangle, \label{eq:alpha2} \\[12pt]
  &(2\bar{g}^{11}\bar{\Gamma}^{3}_{31}-K(r)) \left\langle V \right|\hat{A}_{1}(x^{\mu})\hat{A}^{3}(x'^{\mu})\left| V \right\rangle
   + 2\bar{g}^{33}\bar{\Gamma}^{3}_{23} \left\langle V \right|\hat{A}_{3}(x^{\mu})\hat{A}^{2}(x'^{\mu})\left| V \right\rangle \nonumber \\
  &\quad + (2\bar{g}^{22}\bar{\Gamma}^{3}_{32}-\bar{g}^{33}\bar{\Gamma}^{2}_{33}) \left\langle V \right|\hat{A}_{2}(x^{\mu})\hat{A}^{3}(x'^{\mu})\left| V \right\rangle
   + 2\bar{g}^{33}\bar{\Gamma}^{3}_{13} \left\langle V \right|\hat{A}_{3}(x^{\mu})\hat{A}^{1}(x'^{\mu})\left| V \right\rangle \nonumber \\
  &\qquad = \frac{\eta}{3}(2z-8)\lambda_{0}u_{0}\bar{g}^{33} \left\langle V \right|\hat{A}_{3}(x^{\mu})\hat{A}^{0}(x'^{\mu})\left| V \right\rangle, \label{eq:alpha3}
\end{align}
\end{subequations}
where the function $K(r)$ is defined as:
\begin{equation}
    K(r)=\bar{g}^{00}\bar{\Gamma}^{1}_{00}+\bar{g}^{11}\bar{\Gamma}^{1}_{11}+\bar{g}^{22}\bar{\Gamma}^{1}_{22}+\bar{g}^{33}\bar{\Gamma}^{1}_{33}.
\end{equation}
Once the conditions~(\ref{pep}) are imposed, the cross-correlators $\langle V|\hat{A}_2(x^{\mu})\hat{A}^0(x'^{\mu})|V\rangle$ and $\langle V|\hat{A}_3(x^{\mu})\hat{A}^0(x'^{\mu})|V\rangle$ vanish, and the four equations (\ref{eq:alpha0}), (\ref{eq:alpha1}), (\ref{eq:alpha2}) and (\ref{eq:alpha3}) decouple. From Eq. (\ref{eq:alpha0}), we obtain that
\begin{equation}
\label{A00}
\langle V|\hat{A}_{0}(x^{\mu})\hat{A}^{0}(x'^{\mu})|V\rangle
=
\frac{3a(2z+1)}
{ 2(z-4)\,{\lambda}_{0}}
\langle V|\hat{\phi}(x^{\mu})\hat{\phi}(x'^{\mu})|V\rangle ,
\end{equation}
where we have cancelled the common factor $\eta\, u^0 \bar{g}_{00}$. Furthermore, equations (\ref{eq:alpha1}) and (\ref{eq:alpha2}) must be solved simultaneously for the two angular correlators. Under the condition that both the right-hand side of Eq. (\ref{eq:alpha2}) and the correlator $\langle\hat{A}_1(x^{\mu})\hat{A}^2(x'^{\mu})\rangle$ vanish, substituting the former into Eq. (\ref{eq:alpha1}), we obtain
\begin{equation}
\label{A22}
\langle V|\hat{A}_{2}(x^{\mu})\hat{A}^{2}(x'^{\mu})|V\rangle = -\frac{3a}{{\lambda}_{0}} \bar{g}^{00}\left(\frac{2z+1}{z-4}\right)  \frac{\bar{\Gamma}^{1}_{00}}{\bar{g}^{33}\bar{\Gamma}^{1}_{33}+2\bar{g}^{22}\bar{\Gamma}^{1}_{22}}\langle V|\hat{\phi}(x^{\mu})\hat{\phi}(x'^{\mu})|V\rangle ,
\end{equation}
\begin{equation}
\label{A33}
\langle V|\hat{A}_{3}(x^{\mu})\hat{A}^{3}(x'^{\mu})|V\rangle= -\frac{3a}{2{\lambda}_{0}} \bar{g}^{00}\left(\frac{2z+1}{z-4}\right)  \frac{\bar{\Gamma}^{1}_{00}}{\bar{g}^{33}\bar{\Gamma}^{1}_{33}+2\bar{g}^{22}\bar{\Gamma}^{1}_{22}}\langle V|\hat{\phi}(x^{\mu})\hat{\phi}(x'^{\mu})|V\rangle  .
\end{equation}
After decoupling, the mixed correlator in Eq. (\ref{eq:alpha3}) reduces to
\begin{equation}
\label{A32}
\langle V|\hat{A}_{3}(x^{\mu})\hat{A}^{2}(x'^{\mu})|V\rangle=\left(\frac{\bar{\Gamma}^{2}_{33}}{2\bar{\Gamma}^{3}_{23}}
-\frac{\bar{g}^{22}}{\bar{g}^{33}}\right)\langle V|\hat{A}_{2}(x^{\mu})\hat{A}^{3}(x'^{\mu})|V\rangle .
\end{equation}
The proportionality factor in Eq. (\ref{A32}) is entirely determined by the effective geometry through the metric coefficients and Christoffel symbols. Therefore, the mixed correlator provides a direct measure of how the background curvature influences the angular correlations of the quantized electromagnetic field. Contracting the resolved correlators with $\bar{g}^{\mu\nu}$ and using $\langle V|\hat{A}_1\hat{A}^1|V\rangle = 0$, the invariant scalar reduces to
\begin{equation}
\langle V|\hat{\phi}(x^{\mu},x'^{\mu})|V\rangle = \frac{3a}{2{\lambda}_{0}}\left(\frac{2z+1}{z-4}\right)
\left[1 - \frac{3\,\bar{g}^{00}\,\bar{\Gamma}^{1}_{00}}{\bar{g}^{33}\bar{\Gamma}^{1}_{33}+2\bar{g}^{22}\bar{\Gamma}^{1}_{22}}\right]\langle V|\hat{\phi}(x^{\mu})\hat{\phi}(x'^{\mu})|V\rangle .
\end{equation}
The bracket contains the contribution from the temporal term (the $+1$ term) together with geometry-dependent angular contributions from the Christoffel ratios. For sufficiently large angular contributions, the bracketed factor may vanish or become negative, leading to a change in the sign of $\langle V|\hat{\phi}(x^{\mu},x'^{\mu})|V\rangle$. Such a regime would signal a qualitative modification of the vacuum electromagnetic correlations induced during the evaporation of the BBH.

\begin{figure}[htbp]
    \centering
    \includegraphics[width=0.7\textwidth]{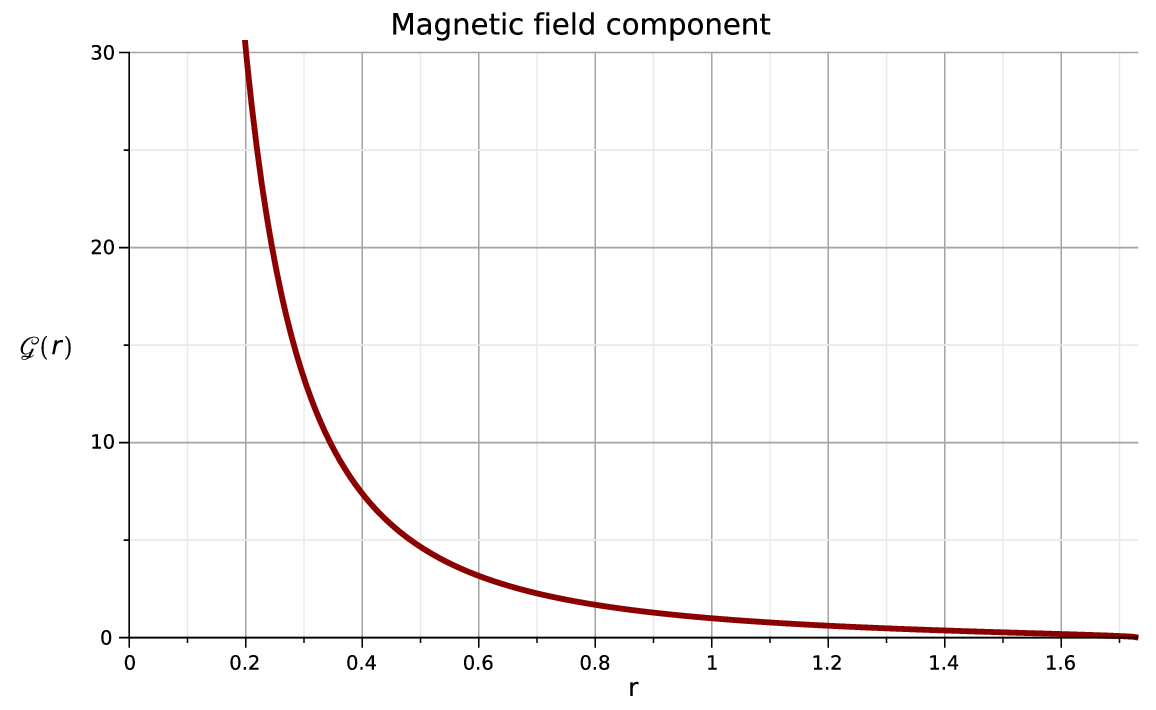}
    \caption{Plot of $\mathcal{G}(r)$ (in units where $\lambda_0=1/m^2$), with $z=2$, $\theta(z)=\dfrac{1+\sqrt5}{2}$.}
    \label{fig:figura1}
\end{figure}

If we use the expression (\ref{B}), we obtain the radial component for the magnetic field. A direct calculation gives
\begin{equation}
\label{br}
\langle V|\hat{B}^1|V\rangle = \frac{u_0}{\sqrt{-\bar{g}}}
\left(2 - \frac{1}{2}\frac{\bar{g}^{33}}{\bar{g}^{22}}
\frac{\bar{\Gamma}^{2}_{33}}{\bar{\Gamma}^{3}_{23}}\right)
\langle V|\hat{A}_2(x^{\mu})\hat{A}_3(x'^{\mu})|V\rangle , \qquad \langle V|\hat{B}^2|V\rangle =0, \qquad \langle V|\hat{B}^3|V\rangle =0,
\end{equation}
which describes magnetic charge, because all the expectation values for the electric field vanish [see equation (\ref{E})]: $\langle V|\hat{E}^i|V\rangle =0$. The explicit calculation for $\langle V|\hat{B}^1|V\rangle$ with the metric in (\ref{gbar1}) and (\ref{gbar2}) gives us the following result for the radial magnetic field in the BBH:

\begin{equation}\label{res_final_g00}
  \langle V|\hat{B}^1|V\rangle
  = \frac{5}{2}\cdot
    \frac{\sqrt{1-\dfrac{\lambda_0 r^2}{3}}}
         {r^2\,\xi^{3/2}(z)}
    \frac{\langle V|\hat{A}_2(x^{\mu})\hat{A}_3(x'^{\mu})|V\rangle}{\sin\vartheta},
\end{equation}
where we have defined
\begin{equation}
\xi(z)= \left[1+\frac{2\left[\theta(z)-1\right]}{z} \right].
\end{equation}
For clarity, we define the prefactor function
\begin{equation}\label{prefactor}
\mathcal{G}(r) \equiv \langle V|\hat{B}^1|V\rangle\,\frac{\sin\vartheta}{\langle V|\hat{A}_2(x^{\mu})\hat{A}_3(x'^{\mu})|V\rangle},
\end{equation}
which isolates the purely geometric, $\vartheta$-independent part of the radial magnetic field. Using this definition, Eq.~(\ref{res_final_g00}) can be written compactly as
\begin{equation}
\mathcal{G}(r) = \frac{5}{2}\cdot
    \frac{\sqrt{1-\dfrac{\lambda_0 r^2}{3}}}
         {r^2\,\xi^{3/2}(z)}.
\end{equation}
The behaviour of $\mathcal{G}(r)$ is illustrated in Fig.~\ref{fig:figura1} as a function of $r$. Notice that the magnetic field is confined to values $r< \sqrt{3/\lambda_0}$.

\subsection{Gauge-field equations and modes normalization }

In order to define the quantum states in the background described by $\bar{g}_{\alpha\beta}$, we must impose the gauge condition in Eq. (\ref{gauge}), which lead to the equations
\begin{equation}\label{gau}
\left<V\right| \hat{\phi}(x^{\beta},x'^{\beta})\left|V\right> =\frac{C}{a} - \frac{\bar{g}^{\mu\nu}}{a} \,\bar{\Gamma}^{\alpha}_{\nu\alpha} \,\left<V\right| \hat{\partial}_{\mu} \hat{\sigma}(x'^{\beta})\left|V\right> =\frac{C}{a} -\frac{b^2}{a} \bar{g}^{\mu\nu}\,\bar{\Gamma}^{\alpha}_{\nu\alpha}\,u_{\mu}\, \left<V\right| \hat{\sigma}(x^{\beta})\hat{\sigma}(x'^{\beta})\left|V\right>,
\end{equation}
for $C$ given by (\ref{CC}). Using the Fourier expansion (\ref{eq:modeexpansion}) with the factorizations (\ref{eq:modefactorization}), and the conditions (\ref{eq:CCRladder}) and (\ref{eq:polcomplete}), we obtain that
\begin{equation}\label{e65}
\left<V\right| \hat{\phi}(x^{\beta},x'^{\beta})\left|V\right> = \left<V\right| \hat{A}_{\mu}(x^{\beta})\,\hat{A}^{\mu}(x'^{\beta})\left|V\right>=
8cb\,\int d^3 {\vec{k}}\,\sqrt{k^2+\bar{\mu}^2}\, \Phi_{\vec{k}}(x'^{\beta})\Phi^*_{\vec{k}}(x^{\beta}),
\end{equation}
where we have used the expression $\omega_{\vec{k}}=c\,\sqrt{k^2+\bar{\mu}^2}$, which is valid for massive photons with mass $\bar{\mu}$. In this framework, for the unit-vectors $\left(\hat{e}_r,\hat{e}_{\vartheta},\hat{e}_{\phi}\right)$, the volume element: $d^3\vec{k}$, represents the volume related to the $3d$-vector $d\vec{k}= \sqrt{-\bar{g}_{11}}\, dk \,\hat{e}_r +
\sqrt{-\bar{g}_{22}}\, k\,d\vartheta \,\hat{e}_{\vartheta} + \sqrt{-\bar{g}_{33}}\, k\,\sin(\vartheta)\, d\phi \,\hat{e}_{\phi}$, which is
\begin{equation}
d^3\vec{k} = \sqrt{-{\rm det}|\bar{g}_{ij}|}\, k^2\,\sin(\vartheta) \,dk\,d\vartheta\,d\phi.
\end{equation}
Furthermore, the analogous expression for $d^3\vec{r}$ is obtained
\begin{equation}
d^3\vec{r} = \sqrt{-{\rm det}|\bar{g}_{ij}|}\, r^2\,\sin(\vartheta) \,dr\,d\vartheta\,d\phi,
\end{equation}
where ${\rm det}|\bar{g}_{ij}|$ denotes the determinant of the submatrix $\bar{g}_{ij}$, for $i,j=1,2,3$. On the other hand, the field $\hat{\sigma}$ can be expressed as
\begin{equation}
    \hat{\sigma}(x^\mu)=\frac{1}{(2\pi)^{3/2}}\,\int d^{3} {\vec{k}}\,
    \left[
        \hat{\sigma}_{\vec{k}}(x^\mu)        +
        \left[\hat{\sigma}_{\vec{k}}(x^\mu)\right]^{\!\dagger}
    \right],
    \label{sigma}
\end{equation}
where $\hat{\sigma}_{\vec{k}}(x^\mu) = \hat{a}_{\vec{k}}\,\varphi_{\vec{k}}(x^{\mu})$ and $\left[\hat{\sigma}_{\vec{k}}(x^\mu)\right]^{\dagger} = \left(\hat{a}_{\vec{k}}\right)^{\dagger} \,\varphi^*_{\vec{k}}(x^{\mu})$. The creation and destruction operators comply with the algebra
\begin{equation}\label{m65}
\left<V\left|\left[\hat{a}_{\vec{k}},\hat{a}^{\dagger}_{{\vec{k}'}}\right]\right|V\right>
=\,\delta^{(3)}\left(\vec{k}-\vec{k'}\right),\quad
\left<V\left|\left[\hat{a}^{\dagger}_{\vec{k}},\hat{a}^{\dagger}_{{\vec{k}'}}\right]\right|V\right>=
\left<V\left|\left[\hat{a}_{\vec{k}},\hat{a}_{{\vec{k}'}}\right]\right|V\right>=0.
\end{equation}
In spherical coordinates the operators $\hat{a}_{\vec{k}}=\hat a_{klm}$ and $\hat{a}^{\dagger}_{\vec{k}}=\hat a_{klm}^{\dagger}$, comply with the algebra
\begin{eqnarray}\label{m56}
&& \left[\hat a_{klm},\;\hat a_{k'l'm'}^{\dagger}\right]= \frac{(2l+1)}{4\pi} \frac{(l-m)!}{(l+m)!} \;\delta ({k}-{k'})\, \delta_{ll'}\,\delta_{mm'},\\
&& \left[\hat a_{klm},\;\hat a_{k'l'm'}\right]=
\left[\hat a_{klm}^{\dagger},\;\hat a_{k'l'm'}^{\dagger}\right]=0,
\end{eqnarray}
for $\hat a_{klm}= \hat a_k\, \sqrt{\frac{(2l+1)}{4\pi} \frac{(l-m)!}{(l+m)!}}$. We can calculate $\left<V\right|\hat{\partial}_{\mu}\hat{\sigma}(x'^\mu)\left|V\right>$, and for ${\vec{k}'}={\vec{b}}$, we obtain
\begin{equation}\label{cc}
u_{\mu}\equiv \left<V\right|\hat{\partial}_{\mu}\hat{\sigma}(x'^{\beta})\left|V\right> = \frac{b^2}{(2\pi)^3}\,u_{\mu}\,\int\,d^3{\vec{k}}\,\varphi_{\vec{k}}(x^{\beta})\varphi^*_{\vec{b}}(x'^{\beta}),
\end{equation}
where $|{\vec{b}}|=b$. This implies the following normalization condition
\begin{equation}\label{normA}
\frac{1}{(2\pi)^3}\,\int\,d^3 {\vec{k}}\,\varphi_{\vec{k}}(x^{\mu})\varphi^*_{\vec{b}}(x^{\mu})=1/b^2,
\end{equation}
for the modes of the gravitational field. Notice that the expectation value for the expression (\ref{quanti}) takes the form
 \begin{equation}
\left<V\right|\left[\hat{ A}_{\nu}(\tau,\vec{r}),\,\frac{\delta {\cal \hat{L}}_{m}}{\delta [\bar{\nabla}_{\beta}\hat{ A}_{\alpha}]}(\tau,\vec{r}\,')\right]\left|V\right>= i\,\hbar \,\delta^{\alpha}_{\,\,\,\nu} \,u^{\beta} \,\delta^{(3)}\left(\vec{r}-\vec{r}\,'\right), \label{quanti1}
\end{equation}
where we have made use of the fact that $u^{\beta}=\langle V | \hat{\sigma}^{\beta} |V\rangle$ [see Eq. (\ref{cc})].

\subsection{Orthogonalization of wave functions in the BBH}

In this section we shall calculate the wave functions $\varphi_{\vec{k}}(x^{\mu})$ and $\Phi_{\vec{k}}(x^{\mu})$ for the gravitational and electromagnetic fields. We must notice that, for the relativistic observer in a time-like solution given by (\ref{u0}), we obtain that $\bar{\Gamma}^{\alpha}_{0\alpha}\,u^{0}=0$, because the are dealing with a static de Sitter metric. Therefore, the equation (\ref{gau}) reduces to
\begin{equation}\label{gaug}
\left<V\right| \hat{\phi}(x^{\beta},x'^{\beta})\left|V\right> =\frac{C}{a} -\frac{b^2}{a} \,\bar{\Gamma}^{\alpha}_{0\alpha}\,u^0\, \left<V\right| \hat{\sigma}(x^{\beta})\hat{\sigma}(x'^{\beta})\left|V\right>=\frac{C}{a},
\end{equation}
such that the result of $\left<V\right| \hat{\phi}(x^{\beta},x'^{\beta})\left|V\right>$ depends on the gauge we choose in (\ref{gauge}). Notice that the particular case where we choose $\lambda_1=0$, gives us $C=0$ and therefore the electromagnetic fluctuations of spacetime vanish: $\left<V\right| \hat{\phi}(x^{\beta},x'^{\beta})\left|V\right>=\left<V\right| \hat{\phi}(x^{\beta},x'^{\beta})\hat{\phi}(x^{\beta},x''^{\beta})\left|V\right>=0$. However, even in this case we recover the magnetic field value due to a magnetic monopole in (\ref{res_final_g00}).

\subsubsection{Orthogonalization of gravitational wave functions}

The equation (\ref{normA}) can be solved by proposing that
\begin{equation}
\varphi_{\vec{k}}(x^{\mu})\varphi^*_{\vec{b}}(x^{\mu}) = \delta^{(3)}\left[b^{2/3}(\vec{k}-\vec{b})\right]=\frac{1}{|b|^2}\, \delta^{(3)}\left(\vec{k}-\vec{b}\right),
\end{equation}
where the exponential representation of $\delta^{(3)}\left(\vec{k}-\vec{b}\right)$ in an arbitrary background metric is
\begin{equation}\label{orto}
\delta^{(3)}\left[(\vec{k}-\vec{b})\right] =\frac{1}{(2\pi)^3} \int d^3\vec{r} \,e^{i\,(\vec{k}-\vec{b})_j\,x^j} \equiv \left<\varphi_{\vec{k}}(x^{\mu})|\varphi_{\vec{b}}(x^{\mu})\right>,
\end{equation}
where $j$ runs on the spatial components and take the values $1,2,3$. Because we are working in spherical coordinates we must set
\begin{equation}
\delta^{(3)}\left[(\vec{k}-\vec{b})\right] = \frac{1}{k^2} \delta(k-b) \delta_{ll'} \,\delta_{mm'}.
\end{equation}
Notice that Eq. (\ref{orto}) gives us the orthogonalization of the spacetime
wave states for
\begin{equation}\label{gravito}
\left<\varphi_{\vec{k}}(x^{\mu})|\vec{r}\right> = \frac{1}{(2\pi)^{3/2}}\,e^{i\,{k}_{j}\,x^{j}}, \quad
\left<\vec{r}\,|\varphi_{\vec{b}}(x^{\mu})\right> = \frac{1}{(2\pi)^{3/2}}\,e^{-i\,{b}_{j}\,x^{j}}.
\end{equation}
In the case of the BBH, the inner products are given by the expressions that take into account the quantum corrections of spacetime
\begin{equation}
{k}_{j}\,x^{j} = \bar{g}_{ij}\,k^{i}\,x^{j}, \qquad {b}_{j}\,x^{j} = \bar{g}_{ij}\,b^{i}\,x^{j},
\end{equation}
such that in spherical coordinates the wavenumber components are $b^{j} \equiv (b,l',m')$ and $k^{i} \equiv (k,l,m)$. Notice that the components of the metric tensor with quantum corrections included are given by the expressions (\ref{gbar1}) and (\ref{gbar2}).

\subsubsection{Orthogonalization of electromagnetic wave functions}

Once we have obtained $\varphi_{\vec{k}}(x^{\mu})$, we can calculate $\Phi_{\vec{k}}(x^{\mu})$ through the expression (\ref{e65}). Using this equation we obtain, for $\omega_{\vec{k}}= \sqrt{k^2+\bar{\mu}^2}$, that
\begin{equation}
\left(\frac{C}{a}\right)\,\delta^{(3)}(\vec{k}-\vec{b}) = \left[[k^2+\bar{\mu}^2]^{1/4}\,\Phi_{\vec{k}}(x^{\mu}) \right]\,\left[[b^2+\bar{\mu}^2]^{1/4}\,\Phi^*_{\vec{b}}(x^{\mu})\right] = \frac{C}{(2\pi)^{3}\,a} \,\int d^3\vec{r}\,e^{i\,(\vec{k}-\vec{b})_jx^j}.
\end{equation}
Therefore, two possible solutions are
\begin{equation}
\left<\Phi_{\vec{k}}(x^{\mu})|\vec{r}\right> = \sqrt{\frac{C}{(2\pi)^{3}\,a}}\,\frac{e^{i\,{k}_{j}\,x^{j}}}{[k^2+\bar{\mu}^2]^{1/4}}, \quad
\left<\vec{r}\,|\Phi_{\vec{b}}(x^{\mu})\right> = \sqrt{\frac{C}{(2\pi)^{3}\,a}}\,\frac{e^{-i\,{b}_{j}\,x^{j}}}{[b^2+\bar{\mu}^2]^{1/4}},
\end{equation}
where $\bar{\mu}$ is the mass of the potential $\hat{A}^{\alpha}$.

\subsection{Lagrangian description of the BBH}

We have shown that the BBH behaves as a massive magnetic monopole that partially evaporates, with the evolution of its mass and magnetic charge being governed by Eqs. (\ref{solu}). Furthermore, we have demonstrated that the only non-vanishing component of the magnetic field is the radial one, which describes a central field confined to distances $r < \sqrt{3/\lambda_0}$. Here, $\lambda_0/\kappa$ represents the vacuum energy density, which depends on the gauge choice $\langle\bar{\nabla}_{\alpha}\hat{\Omega}^{\alpha}\rangle = C$ and is accounted for by a field $\hat{A}^{\alpha}$ with mass $\bar{\mu}$. This means that the BBH Lagrangian density can be expressed through the dual electromagnetic tensor $\hat{\cal{F}}^{\alpha\beta} = \frac{1}{2} \frac{\epsilon^{\alpha\beta\gamma\delta}}{\sqrt{-\bar{g}}}\,F_{\gamma\delta}$, by a generalized Proca
description \cite{proca, proca1,proca2,proca3,proca4,proca5,proca6}
\begin{equation}\label{lag1}
\left<V\right|\hat{\cal{L}}_{m} \left|V\right> = -\frac{\epsilon_0}{4} \, \left<V\right| \hat{\cal{F}}^{\alpha\beta}
\hat{\cal{F}}_{\alpha\beta} \left|V\right> - \left<V\right| \hat{J}^{\alpha}_{(M)}\,\hat{A}_{\alpha} \left|V\right> + \epsilon_0\,\frac{\bar{\mu}^2}{2} \,\left<V\right| \hat{A}^{\alpha}\,\hat{A}_{\alpha} \left|V\right>,
\end{equation}
with a monopole density current
\begin{equation}\label{cu}
\hat{J}^{\alpha}_{(M)}(x^{\mu}) = \eta\epsilon_0\,\left[\frac{2}{3}(z-4)\lambda_0 \,u_{\mu} \,\left<V\right| \bar{\nabla}^{\alpha}\,\hat{A}^{\mu} \left|V\right> - (2z+1)\,a\,\left<V\right| \bar{\nabla}^{\alpha} \left( \hat{A}_{\mu}\,\hat{A}^{\mu}\right) \left|V\right>\right],
\end{equation}
with components $\hat{J}^{\alpha}_{(M)}=\left(c\,\hat{\rho}_{(M)},0,0,0\right)$, such that $\hat{\rho}_{(M)}$ is the monopole density of charge.
The monopole density current (\ref{cu}) is obtained making $ \bar{\mu}^2\epsilon_0\,\left<V\right| \bar{\nabla}_{\mu} \hat{A}^{\mu} \left|V\right>=\left<V\right| \bar{\nabla}_{\mu} \hat{J}^{\mu}_{(M)} \left|V\right> $, in the flux equations (\ref{19}).

Finally, for the action for the BBH on the background metric $\bar{g}_{\alpha\beta}$, could be expressed as
\begin{equation}\label{axion}
\bar{\cal{I}} = \int \,d^4x\, \sqrt{-\bar{g}}\, \left[\frac{1}{2\kappa}\left[\bar{R} + 2\lambda_0\right] + \left<V\right|\hat{\cal{L}}_{m} \left|V\right> \right],
\end{equation}
with $\hat{\cal{L}}_{m}$ given by (\ref{lag1}), the dynamics that describes the system is
\begin{equation}
\left<V\right| \bar{\nabla}_{\beta} \hat{\cal{F}}^{\alpha\beta}
\left|V\right> = \frac{1}{\epsilon_0} \left<V\right| \hat{J}^{\alpha}_{(M)} \left|V\right>.
\end{equation}
This means that the monopole charge density is the gravitoelectromagnetic source of the wave equation for $\hat{A}^{\mu}$ \cite{martin}. Since the magnetic field solution is confined to radii restricted by the vacuum energy, the field $\hat{A}^{\mu}$ describes massive photons (or dark photons) found in the vicinity of a BBH \cite{east,ca,dym,love}.

\section{Conclusions}

By grounding our approach in the conservation of action (which is the fundamental invariant of the system), we have described gravity and electrodynamics as quantum geometric field fluctuations in a partially evaporating BBH with decreasing magnetic charge and mass, such that the remanent background metric is a static de Sitter spacetime, which prevents the BBH from the total evaporation. Regarding the gravitational fluctuation modes of $\langle V|\hat{\sigma}(x^{\mu}) \hat{\sigma}(x'^{\mu})|V\rangle$, they are normalized via Eq. (\ref{normA}) and can be orthogonalized [see Eq. (\ref{gravito})], in agreement with the requirements for a consistent quantization of the gravitational field. On the other hand, the expectation values of the electromagnetic potential components $\hat{A}^{\mu}$ and the gravitational potential $\hat{\sigma}$ are zero, but this is not the case for their partial derivatives, which comply with a nonlinear and noncommutative quantum algebra [see Eqs. (\ref{qa1}), (\ref{qa2}) and (\ref{qa3})]. To calculate
the expectation values of the electric and magnetic fields in the BBH: $\left<V\right|\hat{E}^i\left|V\right>$ and $ \left<V\right|\hat{B}^i\left|V\right>$, we have made use of (\ref{qa1}), (\ref{qa2}) and (\ref{qa3}), in the equations (\ref{dyn}), with the assumptions $ \left<V\right| \hat{A}_{1}(x^{\mu})\hat{A}^{1}(x'^{\mu})\left|V\right>=0$, $\left<V\right| \hat{A}_{1}(x^{\mu})\hat{A}^{i}(x'^{\mu})\left|V\right>=0$ and $\left<V\right| \hat{A}_{0}(x^{\mu})\hat{A}^{i}(x'^{\mu})\left|V\right>=0$. These results describe a compact object with mass and magnetic charge, given in Eq. (\ref{res_final_g00}), where the radial component of the magnetic field is a function of $r$: ${\cal G}(r)$ multiplied by $ \frac{\langle V|\hat{A}_2(x^{\mu})\hat{A}_3(x'^{\mu})|V\rangle}{\sin\vartheta}$. The function ${\cal G}(r)$ was plotted in Fig. (\ref{fig:figura1}) and shows that $\left<V\right| \hat{B}^1\left|V\right>$ is confined to $r<\sqrt{3/\lambda_0}$. An interesting point is that the amplitude of the electromagnetic fluctuations can be fixed by the gauge: $\langle V|\bar{\nabla}_{\alpha} \hat{\Omega}^{\alpha}|V\rangle =C$, in Eq. (\ref{gaug}), which is related to both, the value of $\lambda_0$ and the causal horizon size [which is $0 \leq r<\sqrt{3/\lambda_0}$]. Notice that for the particular case with $C=0$ in  (\ref{CC}), one obtains $\lambda_1=0$ and then the geometric electromagnetic fluctuations: $\left<V\right| \hat{\phi}(x^{\alpha},x'^{\alpha})\left|V\right>=\left<V\right| \hat{A}_{\mu}(x^{\alpha})\hat{A}^{\mu}(x'^{\alpha})\left|V\right>$, vanish. However, even in this case we recover a nonzero radial magnetic field expectation value (\ref{res_final_g00}), which is originated in the correlator $\langle V|\hat{A}_2(x^{\mu})\hat{A}_3(x'^{\mu})|V\rangle$. These results are in correspondence with a massive magnetic monopole in a BBH that radiates massive photons \cite{campa,marriot}. They are stable for $\lambda_0 >0$. In this way, radiating BBHs would be possible candidates for dark matter in the early universe \cite{ko,ca,dym,love}.

Finally, a promising direction for future work would be modeling BBHs with time-dependent vacuum energy to calculate dark matter evolution across various cosmological models with time dependent cosmological parameter \cite{edgar,p2,harko}. While this could be achieved by choosing a time-dependent gauge condition, it remains beyond the scope of this study.

\end{document}